\documentclass[apl,reprint,aps,amsmath,amssymb]{revtex4-1}

\usepackage{graphicx,hyperref,tabularx}

\begin{document}

\title{Resonant dynamics in higher dimensional anti-de Sitter spacetime}

\author{Nils Deppe} \affiliation{Cornell Center for Astrophysics and Planetary
  Science, Cornell University, Ithaca, New York 14853, USA}
\email{nd357@cornell.edu} \date{\today}

\begin{abstract}
  We present results from a detailed study of spherically symmetric
  Einstein-massless-scalar field dynamics with a negative cosmological constant
  in four to nine spacetime dimensions.  This study is the first to present a
  detailed examination of the dynamics in AdS beyond five dimensions, including
  a detailed comparison with numerical solutions of perturbative methods and
  their gauge dependence. Using these perturbative methods, we provide evidence
  that the oscillatory divergence of the first derivative used to argue for
  instability of anti-de Sitter space by Bizo\'n \textit{et al.}~is a
  gauge-dependent effect in five spacetime dimensions but the divergence of the
  second derivative is gauge-independent. We find that the divergence of the
  first derivative appears to be gauge-independent in higher dimensions;
  however, understanding how this divergence depends on the initial data is more
  difficult. We also find that four dimensions is more difficult to study than
  higher dimensions. The results we present show that while much progress has
  been made in understanding the rich dynamics and stability of anti-de Sitter
  space, much work is still to be done. The recent work of Moschidis is
  encouraging that it is possible to understand the problem analytically.
\end{abstract}

\maketitle

\section{Introduction}
Stability of de Sitter and Minkowski spacetimes under small perturbations was
established in 1986\cite{Friedrich1986} and
1993\cite{Christodoulou:1993uv}. Following the Anti-de Sitter (AdS)/Conformal
Field Theory (CFT) conjecture\cite{Maldacena:1997re}, the question of the
stability of AdS became more interesting.  Using the AdS/CFT conjecture it is
possible to address the important question of thermalization and equilibration
of strongly coupled CFTs, which is dual to the question of whether or not small
perturbations of AdS collapse to a black hole.  The stability of AdS against
arbitrarily small scalar field perturbations was first studied numerically in
spherical symmetry\footnote{Novel results beyond spherical symmetry were
  recently presented by Dias and Santos\cite{Dias:2016ewl}.} by Bizo\'n and
Rostworowski in 2011\cite{Bizon:2011gg}, where the authors suggested that a
large class of perturbations eventually collapse to form a black hole even at
arbitrarily small amplitude, $\epsilon$.  However, in such simulations a finite
$\epsilon$ must be used, leaving room for doubt as to whether arbitrarily small
perturbations do actually form a black hole\cite{Dimitrakopoulos:2014ada}.  The
probing of small-amplitude perturbations is aided by the recently proposed
renormalization flow equations (RFEs)\cite{Balasubramanian:2014cja,
  Craps:2014vaa, Craps:2014jwa} for which any behavior observed at amplitude
$\epsilon$ and time $t/\epsilon^2$ is also present at an amplitude $\epsilon'$
and time $t/\epsilon'^2$. This rescaling symmetry was used by Bizo\'n \textit{et
  al.} to argue for the instability of AdS$_5$ based on a divergence in the RFE
solution for specific initial data\cite{Bizon:2015pfa}. However, it is suspected
that this divergence is a gauge-dependent effect\cite{Craps:2015iia}.  The gauge
dependence is understood as an infinite redshift in AdS$_5$ and signals that the
assumption that the system is weakly gravitating is no longer
valid\cite{Dimitrakopoulos:2016tss}. Recently, Moschidis has shown that the
Einstein--null dust system with an inner mirror\cite{Moschidis:2017llu} and the
Einstein--massless Vlasov system\cite{Moschidis:2018ruk} are unstable.

In this paper, we address the AdS stability question and the concerns of
\cite{Craps:2015iia} by performing a detailed study of the RFEs and the
nonlinear Einstein equations.  Our study is the first to examine the gauge
dependence of the RFEs and dynamics in AdS beyond five dimensions.  Our
numerical methods enable us to study the RFEs to a much higher accuracy than
previous work, providing new insight into when the RFEs are no longer valid and
the reasons they fail.  With a new understanding of the RFEs we revisit AdS$_4$,
finding agreement with previous work\cite{Balasubramanian:2014cja,Green:2015dsa}
but strong contrast with what is observed in higher dimensions. Finally, we show
that our results are largely robust against the choice of initial data and
present evidence that the dynamics of AdS$_4$ are more intricate than in higher
dimensions.

\section{Model}
We consider a self-gravitating massless scalar field in a spherically symmetric,
asymptotically AdS spacetime in $d$ spatial dimensions. The metric in
Schwarzschild-like coordinates is
\begin{align}
  \label{eq:metric}
  ds^2=\frac{\ell^2 \left[-Ae^{-2\delta}dt^2+
  A^{-1}dx^2+ \sin^2\left(\frac{x}{\ell}\right)d\Omega^{d-1}\right]}
  {\cos^2\left(\frac{x}{\ell}\right)},
\end{align}
where $d\Omega^{d-1}$ is the metric on $\mathbb{S}^{d-1}$,
$x/\ell\in [0,\pi/2]$, and $t/\ell\in[0,\infty)$.  The areal radius is
$R(x)=\ell\tan(x/\ell)$, and we henceforth work in units of the AdS scale $\ell$
(i.e.~$\ell=1$).

The evolution of the scalar field $\psi$ is governed by the nonlinear system
\begin{align}
  \label{eq:scalarField}
  \Phi_{,t}=\left(Ae^{-\delta}\Pi\right)_{,x},
  \quad\Pi_{,t}=\frac{(Ae^{-\delta}\tan^{d-1}x\Phi)_{,x}}{\tan^{d-1}x},
\end{align}
where $\Pi=A^{-1}e^{\delta}\psi_{,t}$ is the conjugate momentum and
$\Phi=\psi_{,x}$ is an auxiliary variable. The metric functions are solved for
from
\begin{align}
  \label{eq:deltaDeriv}
  \delta_{,x}=&-\sin x\cos x(\Pi^2+\Phi^2)\\
  \label{eq:Aderiv}
  A_{,x}=&\frac{d-2+2\sin^2x}{\sin x\cos x}(1-A)-\sin x\cos x(\Phi^2+\Pi^2).
\end{align}
See \cite{Deppe:2014oua,Deppe:2015qsa} for a detailed discussion of the code we
use to solve this system. At the origin we choose $A(x=0,t)=1$. Two common gauge
choices are the interior time gauge (ITG), where $\delta(x=0,t)=0$, and the
boundary time gauge (BTG), where $\delta(x=\pi/2,t)=0$. We perform evolutions of
the full nonlinear theory in the ITG.

We are particularly interested in perturbations about AdS$_{(d+1)}$ whose
evolution at linear order is governed by
$\hat{L}=-(\tan^{1-d}x)\partial_x(\tan^{d-1}x\partial_x)$ (this can be seen by
setting $A=1,$ $\delta=0$ and $\Pi=\psi_{,t}$ in
Eq.~\eqref{eq:scalarField}). The eigenmodes of $\hat{L}$ are given in terms of
Jacobi polynomials,
\begin{align}
  e_j(x)&=\kappa_j\cos^d(x)P^{(d/2-1,d/2)}_j(\cos 2x)
\end{align}
with eigenvalues $\omega_j=d+2j$ and where
$\kappa_j=2\sqrt{j!(j+d-1)!}/\Gamma(j+d/2)$\cite{Craps:2014jwa}.

Recently much attention has been given to the renormalization flow or two-time
framework
equations\cite{Balasubramanian:2014cja,Craps:2014vaa,Craps:2014jwa,Craps:2015iia,Bizon:2015pfa,Green:2015dsa}. A
detailed study of AdS$_4$ was presented in \cite{Green:2015dsa}, while
\cite{Bizon:2015pfa} investigated AdS$_5$. To study the RFEs, a ``slow time''
$\tau=\epsilon^2t$ is introduced, and dynamics on very short time scales can be
thought of as being averaged over. The scalar field perturbation is expanded as
$\psi(x,t)=\sum_{l=0}^{\infty}A_l\cos(\omega_l t+B_l)e_l(x)$, where $A_l(\tau)$
and $B_l(\tau)$ are time-dependent coefficients. The evolution of $A_l$ and
$B_l$ is given by the RFEs\cite{Craps:2014jwa}
\begin{align}
  \label{eq:amplitude}
  -\frac{dA_l}{d\tau}=&\sum_{\substack{i,j,k\\i+j=k+l\\\{i,j\}\ne\{k,l\}}}
  \frac{S_{ijkl}}{2\omega_l}A_iA_jA_k\sin(B_l+B_k-B_i-B_j),\\
  \label{eq:phase}
  -\frac{dB_l}{d\tau}=&\sum_{\substack{i,j,k\\i+j=k+l\\\{i,j\}\ne\{k,l\}}}
  \frac{S_{ijkl}}{2\omega_lA_l}A_iA_jA_k\cos(B_l+B_k-B_i-B_j)\notag\\
                      &+\frac{T_l}{2\omega_l }A_l^2+\sum_{\substack{i\\i\ne l}}\frac{R_{il}}{2\omega_l}A_i^2,
\end{align}
where $\{i,j\}\ne\{k,l\}$ means both $i$ and $j$ are not equal to $k$ or $l$,
and the coefficients $T_l$, $R_{il}$ and $S_{ijkl}$ are given by integrals over
the eigenmodes in appendix A of \cite{Craps:2014jwa} and by recursion relations
in \cite{Craps:2015iia}.  The gauge dependence of the coefficients is discussed
in \cite{Craps:2015iia}.

In our numerical computations we typically truncate the RFEs
(\ref{eq:amplitude}-\ref{eq:phase}) at $l_{\text{max}}=399$, giving a good
balance between computational cost and accuracy, and refer to this system as the
truncated RFEs (TRFEs). We note that the evolutions dominate the computational
cost, not the construction of $T_l,R_{il}$ and $S_{ijkl}$, which we have
computed to $l_{\rm{max}}>700$. We compute the coefficients $T_l,R_{il}$ and
$S_{ijkl}$ directly by combining the integrals in \cite{Craps:2014jwa}
analytically. For the evaluation we have tested Simpson's rule on a uniform grid
and found this to be computationally inefficient. Ultimately we used adaptive
Gaussian quadrature with a relative error tolerance of $10^{-10}$.

We tested several algorithms for time integration, including the fifth and
eighth order methods by Dormand and Prince, the semi-implicit extrapolation
method, and a fully implicit variable order (orders 5, 9 and 13) Gauss-Radau
method similar to what is used in \cite{Green:2015dsa, Bizon:2015pfa}. We find
that choosing good values for the absolute and relative truncation error is
important since the $A_l$'s decay exponentially and we are interested in their
behavior even early on when they are still small. For our simulations we are
able to conserve the RFE energy, $E=\sum_{l=0}^{l_{\max}}\omega_l{}^2|A_l|^2/4$
to approximately thirteen significant figures and we also see excellent
agreement with numerical evolutions of the nonlinear Einstein equations. The
results presented here used the eighth order Dormand and Prince method.

\section{Results}
We present results from a detailed study of the TRFEs and full nonlinear
numerical evolutions in four to nine spacetime dimensions. For concreteness we
focus on two-mode initial data of the form
\begin{align}
  \label{eq:two-mode}
  \psi(x,0)=\epsilon(e_0(x)+\kappa e_1(x))/d
\end{align}
but have also studied Gaussian initial data given by
\begin{align}
  \label{eq:PiID}
  \Pi(x,0)=\epsilon\exp\left(-\frac{\tan^2(x)}{\sigma^2}\right),
  \;\; \psi(x,0)=0.\
\end{align}
In the evolutions presented here we choose $\kappa=d/(d+2)$, which has been
studied extensively in
AdS$_4$\cite{Balasubramanian:2014cja,Green:2015dsa,Deppe:2015qsa} and in
AdS$_5$\cite{Bizon:2015pfa} using the ITG. A logarithmic divergence in the time
derivative of the phases, $dB_l/d\tau$, was observed in
\cite{Bizon:2015pfa}. This is consistent with an asymptotic analysis of the
equations in the ITG; however, the terms leading to the logarithmic divergence
appear to be absent in the BTG\cite{Craps:2015iia}. We will address this in
detail below.

\begin{figure}[]
  \centering \includegraphics[width=0.47\textwidth]{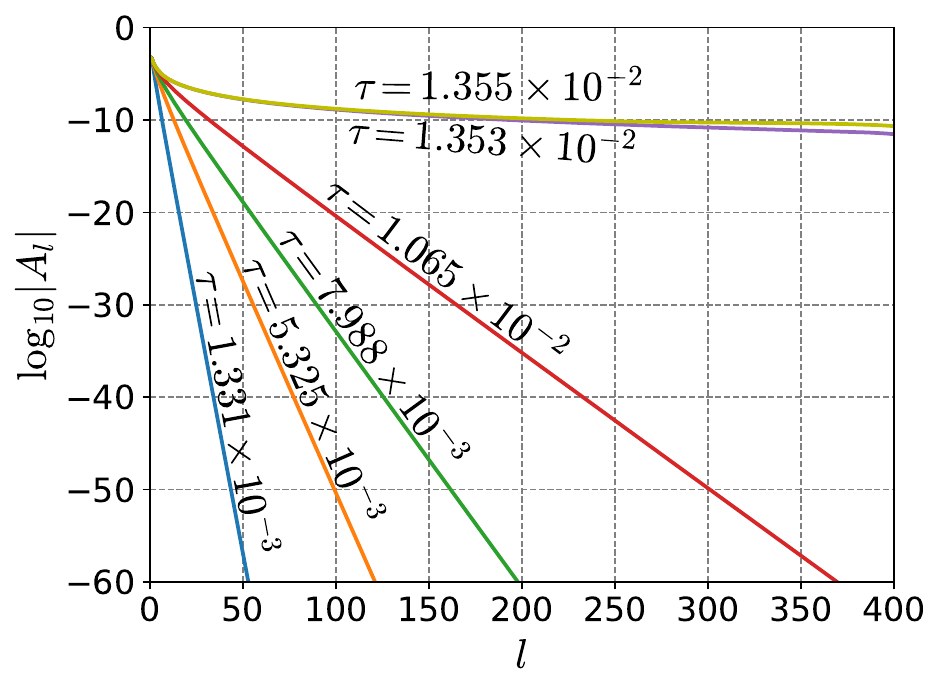}
  \caption{The spectrum $A_l$ with $l_{\text{max}}=399$ for initial
    data~\eqref{eq:two-mode} in AdS$_9$ using the ITG. The spectrum becomes
    singular when $\tau_\star=\epsilon^2t_\star\approx1.353\times10^{-2}$.}
  \label{fig:SpectrumAdS9}
\end{figure}

An interesting technique for analyzing solutions to the TRFEs is the
analyticity-strip method\cite{Sulem1983, Bizon:2015pfa}.  This method involves
fitting the spectrum $A_l$ to
\begin{align}
  \label{eq:analyticityStrip}
  A_l=C(\tau)l^{-\gamma(\tau)}e^{-\rho(\tau)l}
\end{align}
for $l\gg 1$. The analyticity radius $\rho(\tau)$ should be interpreted as the
distance between the real axis and the nearest singularity in the complex
plane\footnote{See Eq.~(2.2) of \cite{Sulem1983} for more details.}. When $\rho$
becomes zero the TRFEs have evolved to a singular spectrum.  We denote the time
when the spectrum becomes singular by $\tau_\star$(or
$t_\star=\tau_\star/\epsilon^2$) and in $d>3$ stop our evolutions of the TRFEs
when $\tau$ is slightly larger than $\tau_\star$. All fits unless otherwise
specified use data from simulations done with $l_{\text{max}}=399$ and omit the
lowest and highest twenty modes to reduce errors from truncation. For
concreteness we present results in AdS$_9$ but observe qualitatively identical
behavior for $d>3$. The spectrum for initial data~\eqref{eq:two-mode} in AdS$_9$
at different times is shown in Fig.~\ref{fig:SpectrumAdS9}.  At
$\tau=1.355\times10^{-2}$ the spectrum is already singular, so we show it only
for completeness.

\begin{figure}[]
  \centering \includegraphics[width=0.47\textwidth]{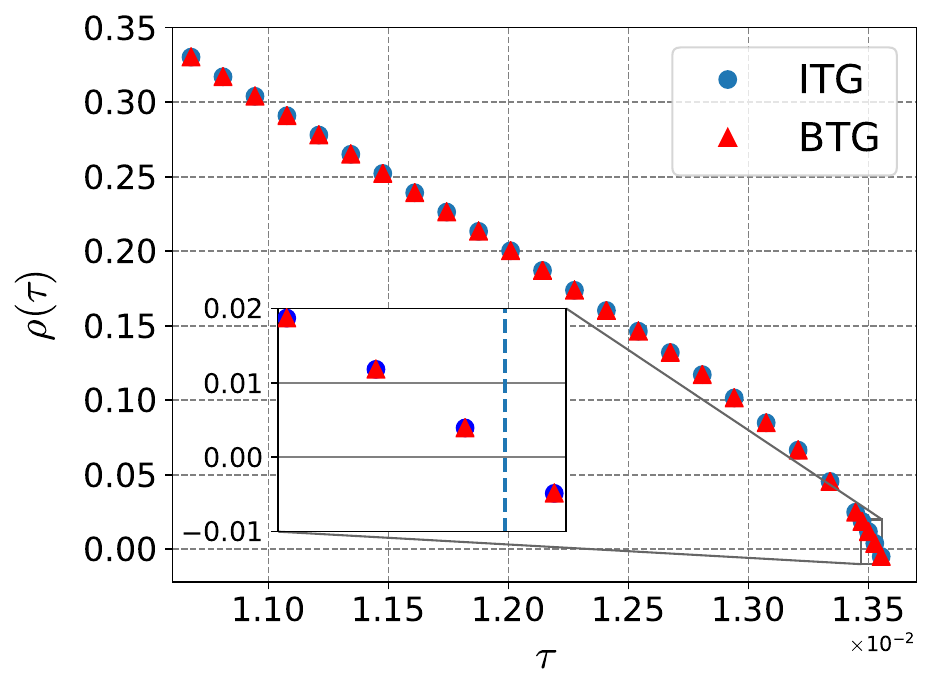}
  \caption{$\rho(\tau)$ for the $l_{\text{max}}=399$ AdS$_9$ evolution in the
    ITG and BTG. In both gauges the spectrum becomes singular at
    $\tau_\star\approx1.353\times10^{-2}$, suggesting this behavior is
    gauge-independent.}
  \label{fig:AnalyticityRadiusAdS9}
\end{figure}

In Fig.~\ref{fig:AnalyticityRadiusAdS9} we plot $\rho(\tau)$ for both the ITG
and BTG for AdS$_9$. We observe that the spectrum becomes singular at
approximately the same $\tau_\star$ in both the BTG and the ITG, independent of
the dimension being studied, suggesting this behavior is
gauge-independent. Interestingly, in our study of $d>3$ we find that the
spectrum becomes singular at approximately the same time that a black hole forms
in the full nonlinear theory, at least for the initial data studied. We will
discuss this further in the context of AdS$_4$ below. Note that the TRFEs should
no longer be trusted when $\rho\le1/l_{\mathrm{max}}$.

Looking at the asymptotic behavior of the $T, R, S$ coefficients for $d>3$ in
the ITG we see that
$S_{\lambda i, \lambda j, \lambda k, \lambda l}\sim \lambda^d S_{ijkl}$,
$T_l\sim l^{d+1}$, and $R_{il}\sim l^2i^{d-1}$\cite{Craps:2015iia}.
Substituting $T_l$, $R_{jl}$, and Eq.~(\ref{eq:analyticityStrip}) into
Eq.~(\ref{eq:phase}) we get
\begin{align}
  -2\omega_l\frac{dB_l}{d\tau}\sim& l^{d+1-2\gamma}e^{-2\rho(\tau_\star-\tau)l}
                                    \notag \\
                                  &+l^2\sum_{\substack{i\\i\ne
  l}}i^{d-1-2\gamma}e^{-2\rho(\tau_\star-\tau)i}.
  \label{eq:deriv_blowup}
\end{align}
Since $\rho\to 0$ as $\tau\to\tau_\star$ the first term goes to a constant as
$\tau\to\tau_\star$. The sum in the second term evaluates to a polylogarithm,
\begin{align}
  \mathrm{Li}_{2\gamma + 1 - d}(e^{-2\rho(\tau_\star-\tau)}).
  \label{eq:polylog_blowup}
\end{align}
In AdS$_{5}$ $\gamma=2$ and we get the results of~\cite{Bizon:2015pfa} that
\begin{align}
  \frac{dB_l}{d\tau}\sim\ln(\tau_\star-\tau).
\end{align}
In the BTG the dominant terms in the coefficients go as
$S_{\lambda i, \lambda j, \lambda k, \lambda l}\sim \lambda^d S_{ijkl}$,
$T_l\sim l^{d}$, and $R_{il}\sim l^2i^{d-2}$. We have verified this by fitting
some of the $R_{il}$ coefficients with $l_{\max}=4100$. Substituting into
Eq.~\eqref{eq:phase} we see that Eq.~\eqref{eq:polylog_blowup} describes the
second derivative of the phases in the BTG while the first derivative goes as
\begin{align}
  \label{eq:deriv blowup BTG}
  -2\omega_l\frac{dB_l}{d\tau}\sim& l^{d-2\gamma}e^{-2\rho(\tau_\star-\tau)l}
                                    \notag \\
                                  &+l^2\sum_{\substack{i\\i\ne
  l}}i^{d-2-2\gamma}e^{-2\rho(\tau_\star-\tau)i} \\
  \sim&\mathrm{Li}_{2\gamma + 2 - d}(e^{-2\rho(\tau_\star-\tau)}).
        \label{eq:polylog blowup BTG}
\end{align}

In AdS$_5$ the asymptotic behavior means that the logarithmic divergence is
present in the second time derivative in the BTG, but the first time derivative
of the phases remains regular.  The presence of the oscillatory singularity in
the ITG but not in the BTG in AdS$_5$ can be understood as the assumption that
the system is weakly gravitating is breaking
down\cite{Dimitrakopoulos:2016tss}. Specifically, the redshift becomes infinite
and so gravity is no longer weak. The oscillatory blowup is related to this
infinite redshift in AdS$_5$, but its nature is not yet understood in higher
dimensions\cite{Dimitrakopoulos:2016tss}. However, the redshift in higher
dimensions does explain why blowup in the time derivative of the phases occurs
at different rates in the ITG and BTG.

\begin{table}
  \centering
  \begin{tabularx}{\columnwidth}{@{\extracolsep{\stretch{1}}}*{6}{c}@{}}
    \hline
    $d$ & Gauge & $\epsilon$ & $\gamma$ & $\tilde{\gamma}$ & $\tau_\star$ \\ \hline
    4 & ITG & 0.0317  & 2 & 2.07 & 0.514 \\
    4 & BTG & 0.0317  & 2 & 2.11 & 0.514 \\
    8 & ITG & 0.00516 & 4 & 3.84 & 0.01354 \\
    8 & BTG & 0.00516 & 4 & 3.54 & 0.01354 \\
    \hline
  \end{tabularx}
  \caption{The estimated values of $\gamma$ and $\tau_\star$ in AdS$_5$ and
    AdS$_9$. We obtain $\gamma$ by fitting Eq.~\eqref{eq:analyticityStrip} at
    $\tau_\star$, and also from fitting Eq.~\eqref{eq:polylog_blowup} and
    Eq.~\eqref{eq:polylog blowup BTG}. The value we obtain from
    Eq.~\ref{eq:polylog_blowup} and Eq.~\ref{eq:polylog blowup BTG} we denote by
    $\tilde{\gamma}$. In the BTG in AdS$_5$ we fit to $d^2B_l/d\tau^2$ instead
    of $dB_l/d\tau$.}\label{tab:gamma dimensions}
\end{table}

\begin{figure}[!ht]
  \centering \includegraphics[width=0.47\textwidth]{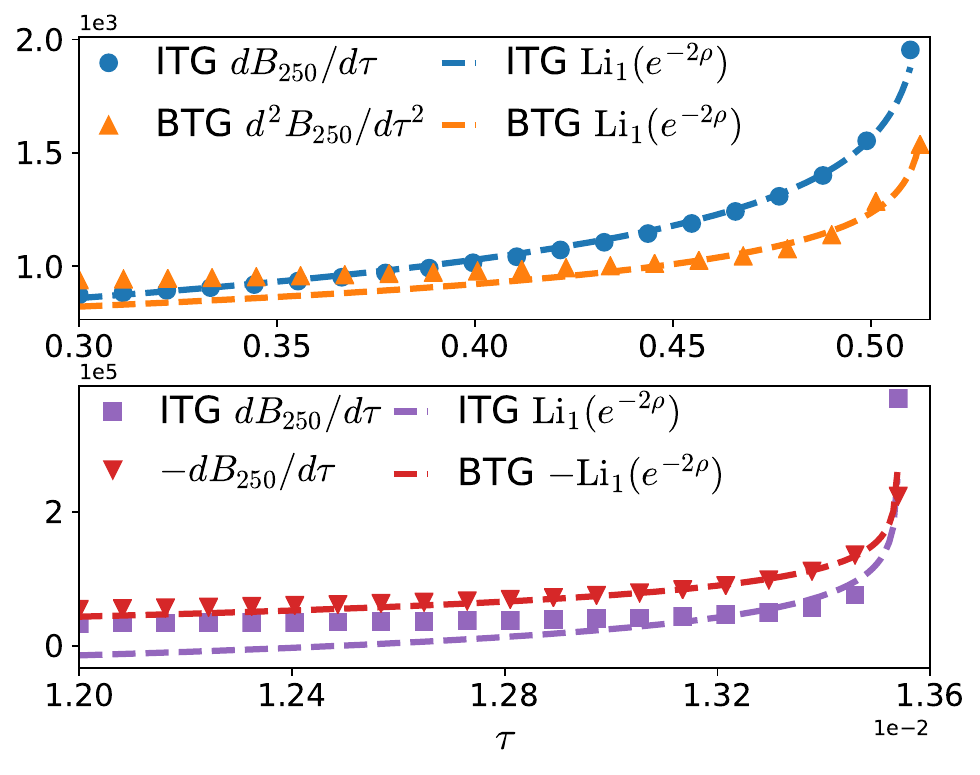}
  \caption{Derivatives of the phase $B_{250}$ in both the ITG and BTG in AdS$_5$
    (top panel) and AdS$_9$ (bottom panel). We show the second derivative of
    $B_l$ in AdS$_5$ in the BTG because the first derivative does not exhibit
    any blowup, as suggested in~\cite{Craps:2015iia}. To assess the behavior we
    fit $\mathrm{Li}_{2\gamma + 1 - d}(e^{-2\rho(\tau_\star-\tau)})$ (dashed
    lines) to the data. Note that in the BTG in AdS$_5$ we rescale the second
    derivative by $1/70$ so that the scales of the first and second derivative
    are comparable.}
  \label{fig:PhaseDeriv}
\end{figure}

Because determining $\gamma$ from the analyticity-strip method is notoriously
difficult we also fit for $\gamma$ keeping $\tau_\star$ fixed, since the time
when the spectrum becomes singular is estimated more robustly from the
analyticity radius. In table~\ref{tab:gamma dimensions} we show results of the
fits in AdS$_5$ and AdS$_9$ in the ITG and BTG, denoting the result of the fit
to Eq.~\eqref{eq:polylog_blowup} in the ITG and Eq.~\eqref{eq:polylog blowup
  BTG} in the BTG by $\tilde{\gamma}$. We note that in the BTG in AdS$_5$ we fit
Eq.~\eqref{eq:polylog_blowup} to $d^2B_l/d\tau^2$. Because of the difficulty of
the fit we only draw the qualitative conclusion that in $d>3$ spatial dimensions
the time derivative of the phases blows up at some finite $\tau_\star$ that
corresponds to black hole formation in the full nonlinear theory, at least in
the cases we have studied. In general the blowup is polylogarithmic, is more
severe in higher dimensions, and is more severe in the ITG than in the BTG.  As
suggested in~\cite{Craps:2015iia}, there is no logarithmic blowup in
$dB_l/d\tau$. However, there is a logarithmic blowup in
$d^2B_l/d\tau^2$\footnote{We are grateful to P.~Bizon, M.~Maliborski, and
  A.~Rostworowski for suggesting we look at how higher derivatives behave.}, and
since the first integral of $\ln(x)$ does not diverge as $x\to0$, $dB_l/d\tau$
does not diverge in the BTG. This is consistent with an asymptotic analysis of
the coefficients. In the top panel of Fig.~\ref{fig:PhaseDeriv} we show
$dB_l/d\tau$ in the ITG and $d^2B_l/d\tau^2$ in the BTG along with logarithmic
fits to the data in AdS$_5$, while in the bottom panel we show $dB_l/d\tau$ in
the ITG and BTG in AdS$_9$. We analyze the $l=250$ mode because it is far below
$l_{\max}$ to minimize errors that stem from mode truncation.

\begin{figure}[!h]
  \centering \includegraphics[width=0.47\textwidth]{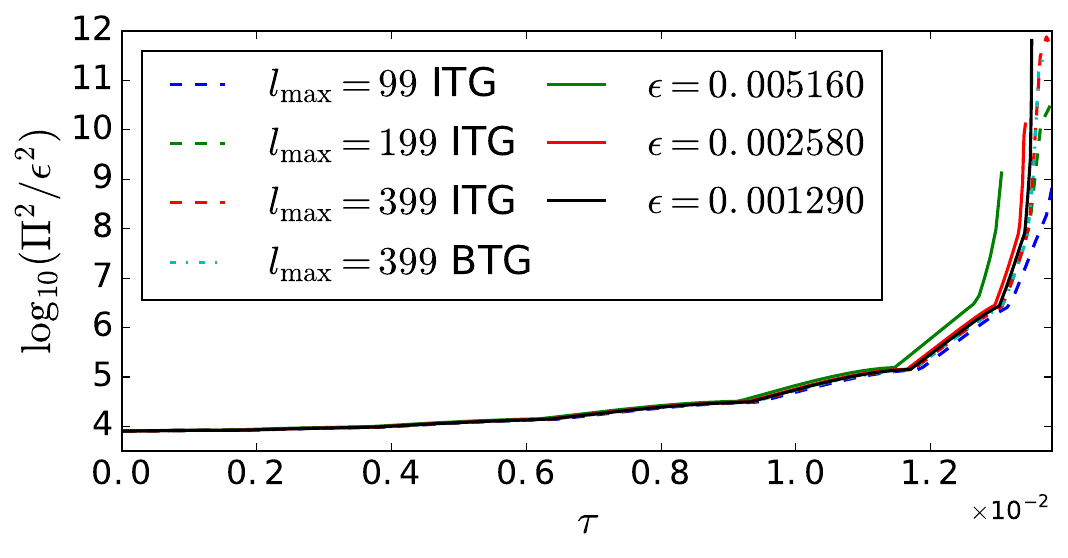}
  \caption{The upper envelope of $\Pi^2(x=0,t)$ for two-mode equal-energy data,
    Eq.~ (\ref{eq:two-mode}), for evolutions in AdS$_9$. Plotted are solutions
    to the TRFEs in the ITG (dash lines) and the BTG (dashed-dotted lines) for
    several different values of $l_{\text{max}}$, and full nonlinear evolutions
    for $\epsilon=0.00516,0.00258,0.00129$ (solid lines). As expected, the
    difference between the TRFE and fully nonlinear solution decreases for
    smaller $\epsilon$ and larger $l_{\rm{max}}$.}
  \label{fig:Pi2AdS9}
\end{figure}

\begin{figure}[!h]
  \centering \includegraphics[width=0.47\textwidth]{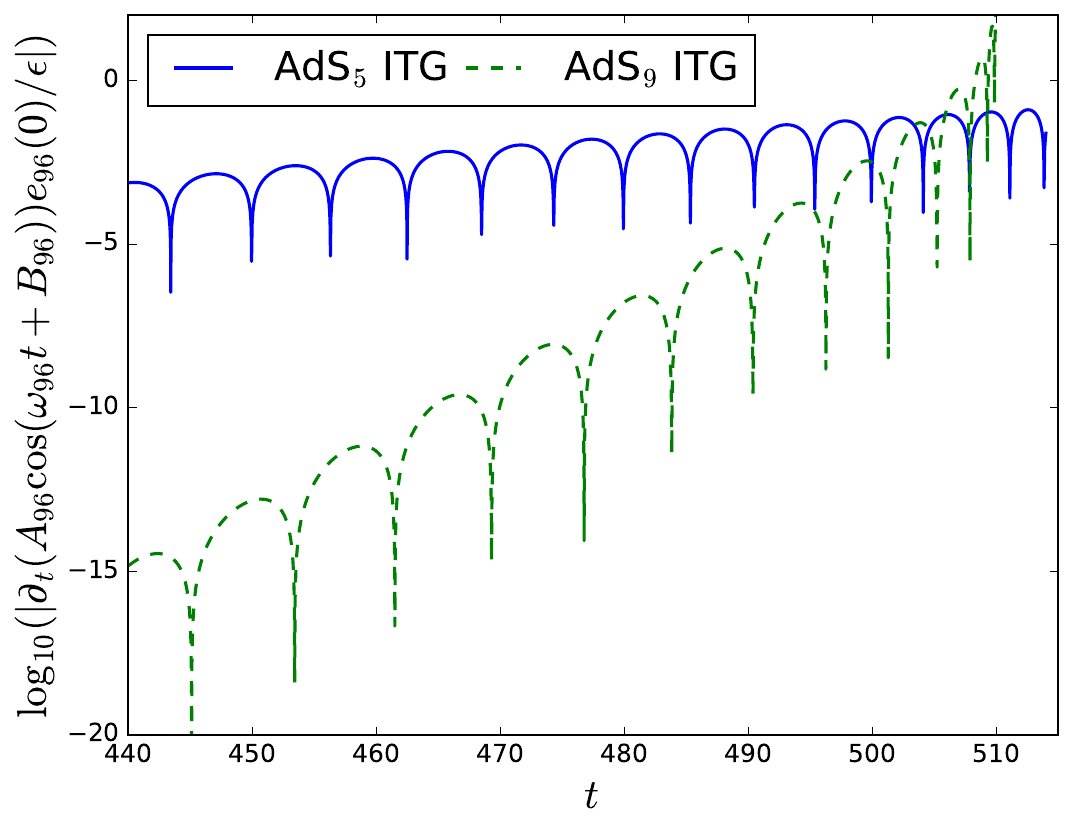}
  \caption{Plot showing how a single term in the series
    $\Pi(x=0,t)=\sum_{l=0}^{l_{\mathrm{max}}}\partial_t(A_l\cos(\omega_l
    t+B_l))e_l(x=0)$ evolves in the ITG comparing AdS$_5$ and AdS$_9$ for
    $l=96$. Similarly behavior is observed for other large $l$ modes and also in
    the BTG.}
  \label{fig:Pi2Comparison}
\end{figure}

In spatial dimensions $d>3$ we observe a direct cascade of energy to higher
modes without any inverse cascades, suggesting the initial data is far from a
quasi-periodic solution\cite{Green:2015dsa}. In Fig.~\ref{fig:Pi2AdS9} we show
the upper envelope of $\Pi^2(x=0,t)/\epsilon^2$, which is proportional to the
Ricci scalar at the origin, for several different values of $l_{\text{max}}$ and
different values of $\epsilon$ for full nonlinear evolutions in AdS$_9$.  There
is good agreement between the fully nonlinear and TRFE solutions and the
agreement improves with increasing dimensionality, at least for
$\Pi^2(x=0,t)/\epsilon^2$. Because
$\Pi(x,t)/\epsilon=\sum_{l=0}^{l_{\mathrm{max}}} \partial_t(A_l\cos(\omega_l
t+B_l))e_l(x=0)/\epsilon$ the improved agreement may be related to the
eigenmodes having larger values at $x=0$ in higher dimensions. In
Fig.~\ref{fig:Pi2Comparison} we plot
$K_l=\partial_t(A_{l}\cos(\omega_{l}t+B_{l}))e_{l}(x=0)/\epsilon$ in the ITG for
$l=96$ in AdS$_5$ and AdS$_9$. We plot $K_l$ as a function of $t$ instead of
$\tau$ so the two simulations are more readily compared. Near the end of the
simulation $K_l$ becomes several orders of magnitude larger in AdS$_9$ than
AdS$_5$. We found qualitatively similar behavior for other values of $l$ and
also in the BTG. However, in the BTG the difference between $K_l$ in AdS$_5$ and
AdS$_9$ is smaller by approximately two orders of magnitude than in the ITG. The
difference in $K_l$ between AdS$_5$ and AdS$_9$ arises mostly from
$e_l(x=0)$. For example, we find that in AdS$_9$ $e_{250}(x=0)$ is $\sim10^4$
times larger than in AdS$_5$.

\begin{figure}[]
  \centering \includegraphics[width=0.47\textwidth]{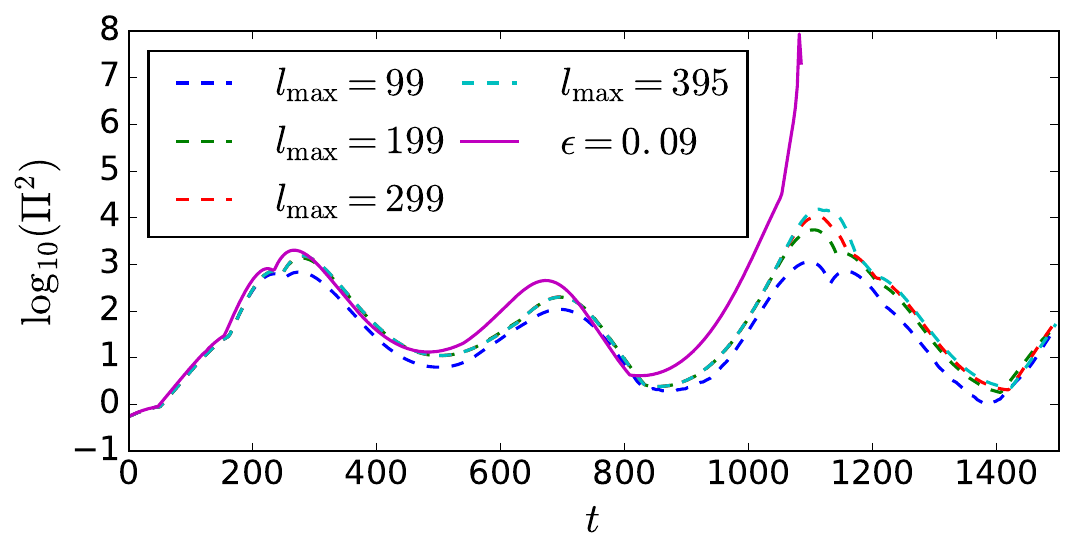}
  \caption{The upper envelope of $\Pi^2(x=0,t)$ for two-mode data
    (\ref{eq:two-mode}) for evolutions in AdS$_4$. Plotted are solutions to the
    TRFEs (dash lines) for different values of $l_{\text{max}}$, and the fully
    nonlinear evolution (solid line) both using the ITG. The TRFE solutions with
    $l_{\text{max}}=299$ and $395$ differ only marginally until the third
    increase in $\Pi^2$.}
  \label{fig:Pi2AdS4}
\end{figure}

We now turn to the case of two-mode equal-energy data, Eq.~\eqref{eq:two-mode},
in AdS$_4$. This case has been studied extensively using numerical relativity
and the TRFEs\cite{Bizon:2011gg, Balasubramanian:2014cja, Bizon:2014bya,
  Balasubramanian:2015uua, Deppe:2015qsa}. It was suggested
in~\cite{Green:2015dsa} that this solution orbits a quasi-periodic solution of
the same temperature. For these evolutions we use $l_{\text{max}}=99,199,299,$
and 395 to test convergence and to understand how the analyticity radius depends
on mode truncation. We compare $\Pi^2(x=0,t)$ with the full nonlinear evolution
in Fig.~\ref{fig:Pi2AdS4}. The 299 and 395 mode evolutions are almost
indistinguishable until the third increase in $\Pi^2$. This suggests that the
agreement between the TRFE and full nonlinear solutions would improve if a lower
amplitude fully nonlinear evolution were studied, similar to what is observed in
Fig.~\ref{fig:Pi2AdS9} for AdS$_9$. Unfortunately, such an evolution requires a
prohibitive amount of computational resources. For concreteness we present
results using the ITG but found the same behavior in the BTG. Because of the
computational expense of using $l_{\mathrm{max}}=395$ we only studied
$l_{\mathrm{max}}=99,199,$ and 299 in the BTG.

\begin{figure}[]
  \centering \includegraphics[width=0.47\textwidth]{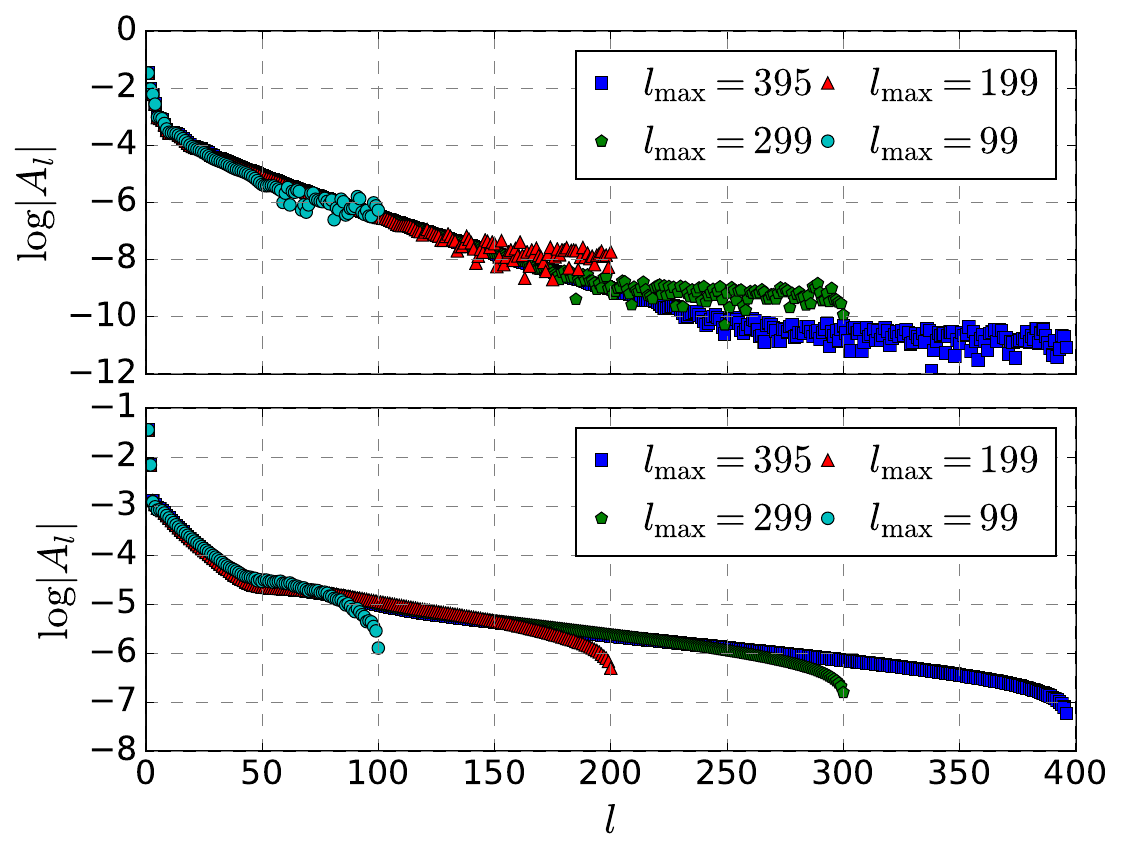}
  \caption{Spectrum of two-mode initial data in AdS$_4$ using the ITG and
    various $l_{\mathrm{max}}$ at $t=1000$ (upper panel) and $t=1100$ (lower
    panel). Mode truncation effects are very apparent at $t=1000$. However, the
    spectrum is better behaved at later times ($t=1100$). Qualitatively similar
    behavior is seen at earlier increases in $\Pi^2$ as well.}
  \label{fig:AdS4twoModeSpectra}
\end{figure}

To better understand the reliability of the TRFEs for the two-mode data in
AdS$_4$ we show the spectrum at two different times\footnote{Here we use $t$
  instead of $\tau$ for easier comparison to the literature.}  $t=1000$ (upper
panel) and $t=1100$ (lower panel) of Fig.~\ref{fig:AdS4twoModeSpectra}.  To gain
an understanding of mode truncation errors and convergence we compare results
using several different values of $l_{\mathrm{max}}$. What we observe is that at
$t=1000$ the highest (in $l$) modes are not well-behaved and that this effect is
dependent on $l_{\mathrm{max}}$, while lower $l$ modes decay
exponentially. Surprisingly, the spectrum using different $l_{\mathrm{max}}$
resemble each other closely again at $t=1100$.  We note that the large $l$ modes
are similarly poorly behaved after the first two increases in $\Pi^2$ in
Fig.~\ref{fig:Pi2AdS4}. What may be surprising is that even with the mode
truncation effects, $\Pi^2$ computed from the TRFEs follows the general trend of
the fully nonlinear evolution quite well.

Because of the agreement in $\Pi^2$ between the TRFE and fully nonlinear
solutions it may be tempting to speculate that the two-mode data in AdS$_4$ is
stable and that the theorems of~\cite{Dimitrakopoulos:2015pwa} for behavior of
stable solutions should be applied for all times. We believe that this would be
premature in light of our new understanding of when the perturbation theory
suffers from mode truncation.  Given the evidence we have presented for when our
results are suffering from mode truncation and the general agreement between
$\Pi^2$ in the fully nonlinear theory and from the TRFEs, the theorems of
\cite{Dimitrakopoulos:2015pwa} suggest that small amplitude equal-energy
two-mode initial data is stable at least until $t\approx1085$ for
$\epsilon=0.09$ ($\tau\approx8.8$). However, we are unable to make predictions
about later time behavior.

\begin{figure}[]
  \centering \includegraphics[width=0.47\textwidth]{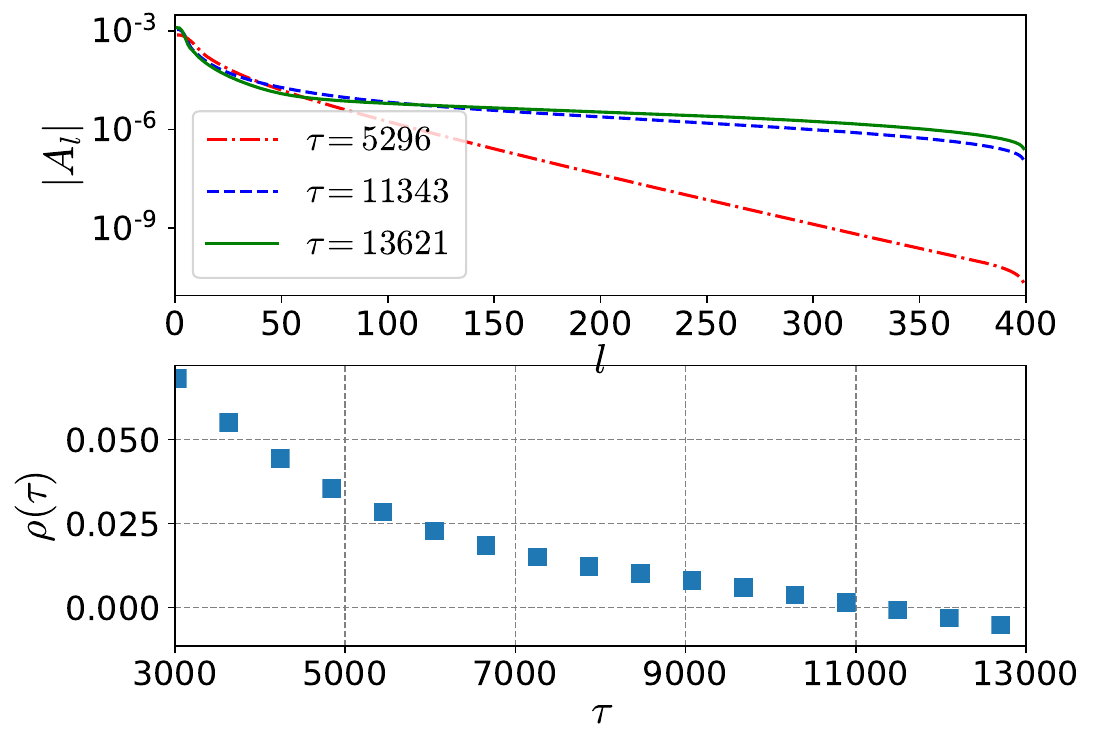}
  \caption{Evolution of Guassian initial data~\eqref{eq:PiID} in AdS$_4$. The
    spectrum is plotted at several times in the upper panel. In the bottom panel
    we plot $\rho(\tau)$ in the ITG fitting Eq.~\eqref{eq:analyticityStrip} to
    the interval $l\in[25, 199]$.}
  \label{fig:AdS4GaussSpectrum}
\end{figure}

Finally, to understand the genericity of our results we also studied initial
data given by Eq.~\eqref{eq:PiID}. Fully nonlinear evolutions of this data have
been well-studied and found to collapse\cite{Bizon:2011gg, Jalmuzna:2011qw,
  Buchel:2012uh, Buchel:2013uba, Deppe:2015qsa}.  We find that the TRFEs require
a larger $l_{\mathrm{max}}$ for Gaussian data than for the two-mode data to
achieve the same accuracy in the sense of how well $\Pi^2(x=0)$ is approximated.
Nevertheless, for $d>3$ we find similar results as for the two-mode data
discussed above.

In the bottom panel of Fig.~\ref{fig:AdS4GaussSpectrum} we plot $\rho(\tau)$,
and in the upper panel we plot $|A_l|$. While testing how $\rho$ depends on the
interval to which Eq.~\eqref{eq:analyticityStrip} is fit we found that if too
many high modes are included then $\rho$ no longer crosses zero, even though a
black hole forms in the full nonlinear theory. While a black hole forming in the
full nonlinear theory does not mean $\rho$ crosses zero, given the sensitivity
to the number of modes to which the fit is done we conclude that in AdS$_4$ many
more modes are needed to accurately follow the dynamics than in higher
dimensions.

\begin{figure}[]
  \centering \includegraphics[width=0.47\textwidth]{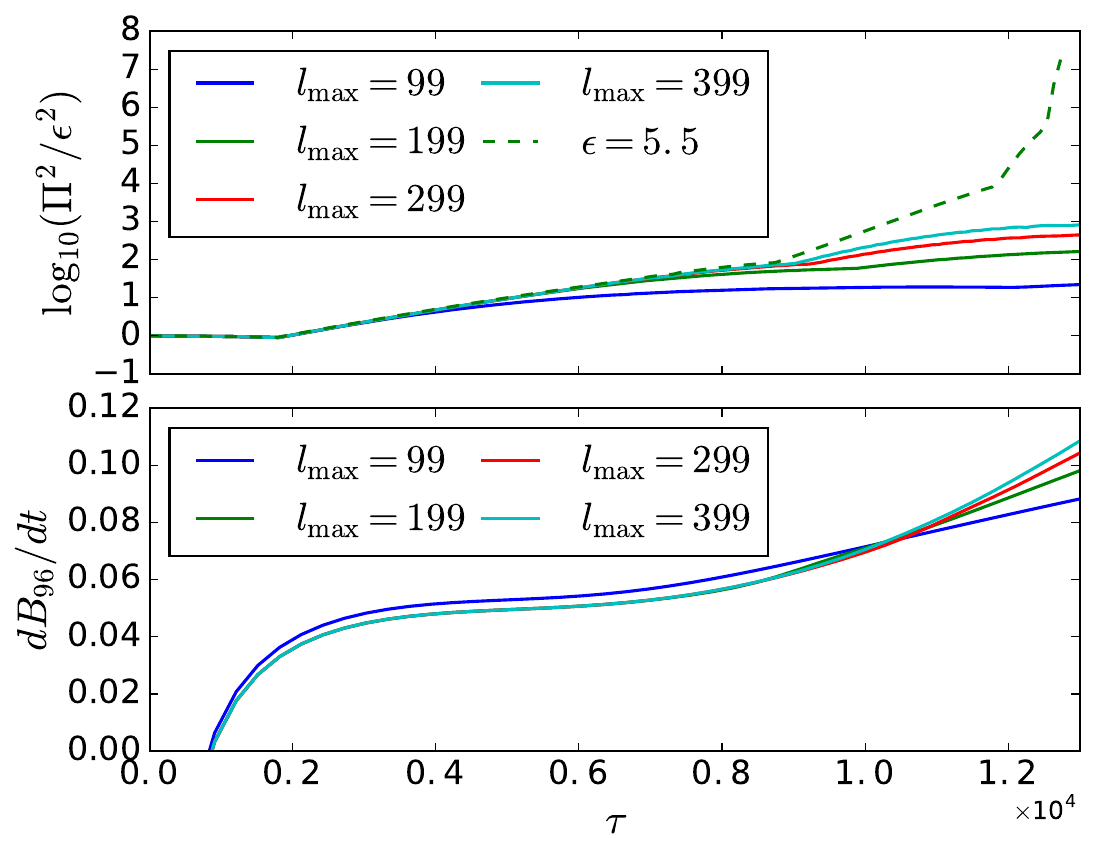}
  \caption{Results from an evolution of Gaussian initial data~\eqref{eq:PiID} in
    AdS$_4$. The upper panel shows the upper envelope of
    $\log_{10}(\Pi^2/\epsilon^2)$ at the origin in the ITG with qualitatively
    identical results in the BTG. The lower panel shows $dB_{96}/dt$ in the ITG
    with similar results for other modes. In the BTG $dB_l/dt$ decreases rather
    than increases with increasing $\tau$ but similarly illustrates that
    $l_{\mathrm{max}}=399$ is insufficient to fully captured the large $l$
    behavior.}
  \label{fig:AdS4Gaussian}
\end{figure}

In the upper panel of Fig.~\ref{fig:AdS4Gaussian} we plot $\Pi^2/\epsilon^2$
showing that $l_{\mathrm{max}}=399$ does not capture the behavior nearly as well
as it does for two-mode data and in higher dimensions (compare to
Fig.~\ref{fig:Pi2AdS9}). We plot $dB_{96}/dt$ for $l_{\mathrm{max}}=99,199,299$
and 399 in the lower panel of Fig.~\ref{fig:AdS4Gaussian}.  In agreement with
\cite{Green:2015dsa}, we do not observe a divergence in $dB_l/dt$ in
AdS$_4$. However, we cannot rule out that such behavior does not appear if
$l_{\mathrm{max}}\gg400$ is used or that higher derivatives also do not
diverge. These findings suggest that at the TRFEs are much more sensitive to
mode trunction in AdS$_4$ than in higher dimensions.

\section{Conclusion}

In summary, our study is the first to examine the gauge dependence of the RFEs
and dynamics in AdS beyond five dimensions. Our numerical methods allow us to
test the RFEs to a much higher accuracy than previous studies, providing new
insight into when the equations no longer accurately approximate the Einstein
equations.  We provide evidence that the oscillatory singularity of the RFEs in
the first derivative used to argue for the instability of AdS$_5$
in~\cite{Bizon:2015pfa} is a gauge-dependent effect in five dimensions and that
this behavior is independent of initial data. However, in the BTG the second
derivative of the phases diverges. Additionally, the divergence of the first
derivative of the phases appears to be gauge-independent for dimensions greater
than five. In agreement with~\cite{Bizon:2015pfa} we find that in $d>3$ the
singular behavior of the RFEs, i.e.~the spectrum becomes singular, occurs at
approximately the same time that a black hole forms in the full nonlinear
theory.  We find that the TRFEs approximate the full nonlinear theory very well
and that in AdS$_5$ through AdS$_9$ the primary source of discrepancy between
the TRFEs and the full nonlinear theory is from mode truncation. However,
AdS$_4$ proves significantly more difficult to study than higher dimensions. We
find that in AdS$_4$ the TRFEs require many more modes in order to accurately
follow the dynamics in regimes where a black hole forms in the full nonlinear
theory, such as the Gaussian initial data (compare the upper panel of
Fig.~\ref{fig:AdS4Gaussian} to Fig.~\ref{fig:Pi2AdS9}). While our results aid in
understanding the validity and behavior of the RFEs, they also show that even
though much progress has been made in understanding the (in)stability of AdS,
there is still much work to be done. Recent work by
Moschidis~\cite{Moschidis:2017llu,Moschidis:2018ruk} is encouraging that it is
possible to understand the AdS (in)stability problem analytically.

\section{Acknowledgements}

We are grateful to Andy Bohn, Brad Cownden, Andrew Frey, Fran\c{c}ois
H\'{e}bert, Lawrence Kidder and Saul Teukolsky for insightful discussions and
feedback on earlier versions of this manuscript. We are also grateful to
P.~Bizon, M.~Maliborski and A.~Rostworowski for insightful discussions.  This
work was supported in part by a Natural Sciences and Engineering Research
Council of Canada PGS-D grant to ND, NSF Grants PHY-1606654 at Cornell
University, and by a grant from the Sherman Fairchild Foundation. Computations
were enabled in part by support provided by WestGrid (www.westgrid.ca) and
Compute Canada Calcul Canada (www.computecanada.ca). Computations were also
performed on the Zwicky cluster at Caltech, supported by the Sherman Fairchild
Foundation and by NSF award PHY-0960291.

\bibliography{refs}

\begin{thebibliography}{28}%
\makeatletter
\providecommand \@ifxundefined [1]{%
 \@ifx{#1\undefined}
}%
\providecommand \@ifnum [1]{%
 \ifnum #1\expandafter \@firstoftwo
 \else \expandafter \@secondoftwo
 \fi
}%
\providecommand \@ifx [1]{%
 \ifx #1\expandafter \@firstoftwo
 \else \expandafter \@secondoftwo
 \fi
}%
\providecommand \natexlab [1]{#1}%
\providecommand \enquote  [1]{``#1''}%
\providecommand \bibnamefont  [1]{#1}%
\providecommand \bibfnamefont [1]{#1}%
\providecommand \citenamefont [1]{#1}%
\providecommand \href@noop [0]{\@secondoftwo}%
\providecommand \href [0]{\begingroup \@sanitize@url \@href}%
\providecommand \@href[1]{\@@startlink{#1}\@@href}%
\providecommand \@@href[1]{\endgroup#1\@@endlink}%
\providecommand \@sanitize@url [0]{\catcode `\\12\catcode `\$12\catcode
  `\&12\catcode `\#12\catcode `\^12\catcode `\_12\catcode `\%12\relax}%
\providecommand \@@startlink[1]{}%
\providecommand \@@endlink[0]{}%
\providecommand \url  [0]{\begingroup\@sanitize@url \@url }%
\providecommand \@url [1]{\endgroup\@href {#1}{\urlprefix }}%
\providecommand \urlprefix  [0]{URL }%
\providecommand \Eprint [0]{\href }%
\providecommand \doibase [0]{http://dx.doi.org/}%
\providecommand \selectlanguage [0]{\@gobble}%
\providecommand \bibinfo  [0]{\@secondoftwo}%
\providecommand \bibfield  [0]{\@secondoftwo}%
\providecommand \translation [1]{[#1]}%
\providecommand \BibitemOpen [0]{}%
\providecommand \bibitemStop [0]{}%
\providecommand \bibitemNoStop [0]{.\EOS\space}%
\providecommand \EOS [0]{\spacefactor3000\relax}%
\providecommand \BibitemShut  [1]{\csname bibitem#1\endcsname}%
\let\auto@bib@innerbib\@empty
\bibitem [{\citenamefont {Friedrich}(1986)}]{Friedrich1986}%
  \BibitemOpen
  \bibfield  {author} {\bibinfo {author} {\bibfnamefont {H.}~\bibnamefont
  {Friedrich}},\ }\href {\doibase 10.1016/0393-0440(86)90004-5} {\bibfield
  {journal} {\bibinfo  {journal} {Journal of Geometry and Physics}\ }\textbf
  {\bibinfo {volume} {3}},\ \bibinfo {pages} {101} (\bibinfo {year}
  {1986})}\BibitemShut {NoStop}%
\bibitem [{\citenamefont {Christodoulou}\ and\ \citenamefont
  {Klainerman}(2014)}]{Christodoulou:1993uv}%
  \BibitemOpen
  \bibfield  {author} {\bibinfo {author} {\bibfnamefont {D.}~\bibnamefont
  {Christodoulou}}\ and\ \bibinfo {author} {\bibfnamefont {S.}~\bibnamefont
  {Klainerman}},\ }\href@noop {} {\emph {\bibinfo {title} {The Global Nonlinear
  Stability of the Minkowski Space (PMS-41)}}},\ Princeton Legacy Library\
  (\bibinfo  {publisher} {Princeton University Press},\ \bibinfo {year}
  {2014})\BibitemShut {NoStop}%
\bibitem [{\citenamefont {Maldacena}(1999)}]{Maldacena:1997re}%
  \BibitemOpen
  \bibfield  {author} {\bibinfo {author} {\bibfnamefont {J.~M.}\ \bibnamefont
  {Maldacena}},\ }\href {\doibase 10.1023/A:1026654312961} {\bibfield
  {journal} {\bibinfo  {journal} {Int. J. Theor. Phys.}\ }\textbf {\bibinfo
  {volume} {38}},\ \bibinfo {pages} {1113} (\bibinfo {year} {1999})},\ \bibinfo
  {note} {[Adv. Theor. Math. Phys.2,231(1998)]},\ \Eprint
  {http://arxiv.org/abs/hep-th/9711200} {arXiv:hep-th/9711200 [hep-th]}
  \BibitemShut {NoStop}%
\bibitem [{Note1()}]{Note1}%
  \BibitemOpen
  \bibinfo {note} {Novel results beyond spherical symmetry were recently
  presented by Dias and Santos\cite {Dias:2016ewl}.}\BibitemShut {Stop}%
\bibitem [{\citenamefont {Bizo\'{n}}\ and\ \citenamefont
  {Rostworowski}(2011)}]{Bizon:2011gg}%
  \BibitemOpen
  \bibfield  {author} {\bibinfo {author} {\bibfnamefont {P.}~\bibnamefont
  {Bizo\'{n}}}\ and\ \bibinfo {author} {\bibfnamefont {A.}~\bibnamefont
  {Rostworowski}},\ }\href {\doibase 10.1103/PhysRevLett.107.031102} {\bibfield
   {journal} {\bibinfo  {journal} {Phys. Rev. Lett.}\ }\textbf {\bibinfo
  {volume} {107}},\ \bibinfo {pages} {031102} (\bibinfo {year} {2011})},\
  \Eprint {http://arxiv.org/abs/1104.3702} {arXiv:1104.3702 [gr-qc]}
  \BibitemShut {NoStop}%
\bibitem [{\citenamefont {Dimitrakopoulos}\ \emph {et~al.}(2015)\citenamefont
  {Dimitrakopoulos}, \citenamefont {Freivogel}, \citenamefont {Lippert},\ and\
  \citenamefont {Yang}}]{Dimitrakopoulos:2014ada}%
  \BibitemOpen
  \bibfield  {author} {\bibinfo {author} {\bibfnamefont {F.~V.}\ \bibnamefont
  {Dimitrakopoulos}}, \bibinfo {author} {\bibfnamefont {B.}~\bibnamefont
  {Freivogel}}, \bibinfo {author} {\bibfnamefont {M.}~\bibnamefont {Lippert}},
  \ and\ \bibinfo {author} {\bibfnamefont {I.-S.}\ \bibnamefont {Yang}},\
  }\href {\doibase 10.1007/JHEP08(2015)077} {\bibfield  {journal} {\bibinfo
  {journal} {JHEP}\ }\textbf {\bibinfo {volume} {08}},\ \bibinfo {pages} {077}
  (\bibinfo {year} {2015})},\ \Eprint {http://arxiv.org/abs/1410.1880}
  {arXiv:1410.1880 [hep-th]} \BibitemShut {NoStop}%
\bibitem [{\citenamefont {Balasubramanian}\ \emph {et~al.}(2014)\citenamefont
  {Balasubramanian}, \citenamefont {Buchel}, \citenamefont {Green},
  \citenamefont {Lehner},\ and\ \citenamefont
  {Liebling}}]{Balasubramanian:2014cja}%
  \BibitemOpen
  \bibfield  {author} {\bibinfo {author} {\bibfnamefont {V.}~\bibnamefont
  {Balasubramanian}}, \bibinfo {author} {\bibfnamefont {A.}~\bibnamefont
  {Buchel}}, \bibinfo {author} {\bibfnamefont {S.~R.}\ \bibnamefont {Green}},
  \bibinfo {author} {\bibfnamefont {L.}~\bibnamefont {Lehner}}, \ and\ \bibinfo
  {author} {\bibfnamefont {S.~L.}\ \bibnamefont {Liebling}},\ }\href {\doibase
  10.1103/PhysRevLett.113.071601} {\bibfield  {journal} {\bibinfo  {journal}
  {Phys. Rev. Lett.}\ }\textbf {\bibinfo {volume} {113}},\ \bibinfo {pages}
  {071601} (\bibinfo {year} {2014})},\ \Eprint {http://arxiv.org/abs/1403.6471}
  {arXiv:1403.6471 [hep-th]} \BibitemShut {NoStop}%
\bibitem [{\citenamefont {Craps}\ \emph {et~al.}(2014)\citenamefont {Craps},
  \citenamefont {Evnin},\ and\ \citenamefont {Vanhoof}}]{Craps:2014vaa}%
  \BibitemOpen
  \bibfield  {author} {\bibinfo {author} {\bibfnamefont {B.}~\bibnamefont
  {Craps}}, \bibinfo {author} {\bibfnamefont {O.}~\bibnamefont {Evnin}}, \ and\
  \bibinfo {author} {\bibfnamefont {J.}~\bibnamefont {Vanhoof}},\ }\href
  {\doibase 10.1007/JHEP10(2014)048} {\bibfield  {journal} {\bibinfo  {journal}
  {JHEP}\ }\textbf {\bibinfo {volume} {10}},\ \bibinfo {pages} {048} (\bibinfo
  {year} {2014})},\ \Eprint {http://arxiv.org/abs/1407.6273} {arXiv:1407.6273
  [gr-qc]} \BibitemShut {NoStop}%
\bibitem [{\citenamefont {Craps}\ \emph
  {et~al.}(2015{\natexlab{a}})\citenamefont {Craps}, \citenamefont {Evnin},\
  and\ \citenamefont {Vanhoof}}]{Craps:2014jwa}%
  \BibitemOpen
  \bibfield  {author} {\bibinfo {author} {\bibfnamefont {B.}~\bibnamefont
  {Craps}}, \bibinfo {author} {\bibfnamefont {O.}~\bibnamefont {Evnin}}, \ and\
  \bibinfo {author} {\bibfnamefont {J.}~\bibnamefont {Vanhoof}},\ }\href
  {\doibase 10.1007/JHEP01(2015)108} {\bibfield  {journal} {\bibinfo  {journal}
  {JHEP}\ }\textbf {\bibinfo {volume} {01}},\ \bibinfo {pages} {108} (\bibinfo
  {year} {2015}{\natexlab{a}})},\ \Eprint {http://arxiv.org/abs/1412.3249}
  {arXiv:1412.3249 [gr-qc]} \BibitemShut {NoStop}%
\bibitem [{\citenamefont {Bizo\'{n}}\ \emph {et~al.}(2015)\citenamefont
  {Bizo\'{n}}, \citenamefont {Maliborski},\ and\ \citenamefont
  {Rostworowski}}]{Bizon:2015pfa}%
  \BibitemOpen
  \bibfield  {author} {\bibinfo {author} {\bibfnamefont {P.}~\bibnamefont
  {Bizo\'{n}}}, \bibinfo {author} {\bibfnamefont {M.}~\bibnamefont
  {Maliborski}}, \ and\ \bibinfo {author} {\bibfnamefont {A.}~\bibnamefont
  {Rostworowski}},\ }\href {\doibase 10.1103/PhysRevLett.115.081103} {\bibfield
   {journal} {\bibinfo  {journal} {Phys. Rev. Lett.}\ }\textbf {\bibinfo
  {volume} {115}},\ \bibinfo {pages} {081103} (\bibinfo {year} {2015})},\
  \Eprint {http://arxiv.org/abs/1506.03519} {arXiv:1506.03519 [gr-qc]}
  \BibitemShut {NoStop}%
\bibitem [{\citenamefont {Craps}\ \emph
  {et~al.}(2015{\natexlab{b}})\citenamefont {Craps}, \citenamefont {Evnin},\
  and\ \citenamefont {Vanhoof}}]{Craps:2015iia}%
  \BibitemOpen
  \bibfield  {author} {\bibinfo {author} {\bibfnamefont {B.}~\bibnamefont
  {Craps}}, \bibinfo {author} {\bibfnamefont {O.}~\bibnamefont {Evnin}}, \ and\
  \bibinfo {author} {\bibfnamefont {J.}~\bibnamefont {Vanhoof}},\ }\href
  {\doibase 10.1007/JHEP10(2015)079} {\bibfield  {journal} {\bibinfo  {journal}
  {JHEP}\ }\textbf {\bibinfo {volume} {10}},\ \bibinfo {pages} {079} (\bibinfo
  {year} {2015}{\natexlab{b}})},\ \Eprint {http://arxiv.org/abs/1508.04943}
  {arXiv:1508.04943 [gr-qc]} \BibitemShut {NoStop}%
\bibitem [{\citenamefont {Dimitrakopoulos}\ \emph {et~al.}(2016)\citenamefont
  {Dimitrakopoulos}, \citenamefont {Freivogel}, \citenamefont {Pedraza},\ and\
  \citenamefont {Yang}}]{Dimitrakopoulos:2016tss}%
  \BibitemOpen
  \bibfield  {author} {\bibinfo {author} {\bibfnamefont {F.~V.}\ \bibnamefont
  {Dimitrakopoulos}}, \bibinfo {author} {\bibfnamefont {B.}~\bibnamefont
  {Freivogel}}, \bibinfo {author} {\bibfnamefont {J.~F.}\ \bibnamefont
  {Pedraza}}, \ and\ \bibinfo {author} {\bibfnamefont {I.-S.}\ \bibnamefont
  {Yang}},\ }\href {\doibase 10.1103/PhysRevD.94.124008} {\bibfield  {journal}
  {\bibinfo  {journal} {Phys. Rev.}\ }\textbf {\bibinfo {volume} {D94}},\
  \bibinfo {pages} {124008} (\bibinfo {year} {2016})},\ \Eprint
  {http://arxiv.org/abs/1607.08094} {arXiv:1607.08094 [hep-th]} \BibitemShut
  {NoStop}%
\bibitem [{\citenamefont {Green}\ \emph {et~al.}(2015)\citenamefont {Green},
  \citenamefont {Maillard}, \citenamefont {Lehner},\ and\ \citenamefont
  {Liebling}}]{Green:2015dsa}%
  \BibitemOpen
  \bibfield  {author} {\bibinfo {author} {\bibfnamefont {S.~R.}\ \bibnamefont
  {Green}}, \bibinfo {author} {\bibfnamefont {A.}~\bibnamefont {Maillard}},
  \bibinfo {author} {\bibfnamefont {L.}~\bibnamefont {Lehner}}, \ and\ \bibinfo
  {author} {\bibfnamefont {S.~L.}\ \bibnamefont {Liebling}},\ }\href {\doibase
  10.1103/PhysRevD.92.084001} {\bibfield  {journal} {\bibinfo  {journal} {Phys.
  Rev.}\ }\textbf {\bibinfo {volume} {D92}},\ \bibinfo {pages} {084001}
  (\bibinfo {year} {2015})},\ \Eprint {http://arxiv.org/abs/1507.08261}
  {arXiv:1507.08261 [gr-qc]} \BibitemShut {NoStop}%
\bibitem [{\citenamefont {Deppe}\ \emph {et~al.}(2015)\citenamefont {Deppe},
  \citenamefont {Kolly}, \citenamefont {Frey},\ and\ \citenamefont
  {Kunstatter}}]{Deppe:2014oua}%
  \BibitemOpen
  \bibfield  {author} {\bibinfo {author} {\bibfnamefont {N.}~\bibnamefont
  {Deppe}}, \bibinfo {author} {\bibfnamefont {A.}~\bibnamefont {Kolly}},
  \bibinfo {author} {\bibfnamefont {A.}~\bibnamefont {Frey}}, \ and\ \bibinfo
  {author} {\bibfnamefont {G.}~\bibnamefont {Kunstatter}},\ }\href {\doibase
  10.1103/PhysRevLett.114.071102} {\bibfield  {journal} {\bibinfo  {journal}
  {Phys. Rev. Lett.}\ }\textbf {\bibinfo {volume} {114}},\ \bibinfo {pages}
  {071102} (\bibinfo {year} {2015})},\ \Eprint {http://arxiv.org/abs/1410.1869}
  {arXiv:1410.1869 [hep-th]} \BibitemShut {NoStop}%
\bibitem [{\citenamefont {Deppe}\ and\ \citenamefont
  {Frey}(2015)}]{Deppe:2015qsa}%
  \BibitemOpen
  \bibfield  {author} {\bibinfo {author} {\bibfnamefont {N.}~\bibnamefont
  {Deppe}}\ and\ \bibinfo {author} {\bibfnamefont {A.~R.}\ \bibnamefont
  {Frey}},\ }\href {\doibase 10.1007/JHEP12(2015)004} {\bibfield  {journal}
  {\bibinfo  {journal} {JHEP}\ }\textbf {\bibinfo {volume} {12}},\ \bibinfo
  {pages} {004} (\bibinfo {year} {2015})},\ \Eprint
  {http://arxiv.org/abs/1508.02709} {arXiv:1508.02709 [hep-th]} \BibitemShut
  {NoStop}%
\bibitem [{\citenamefont {Sulem}\ \emph {et~al.}(1983)\citenamefont {Sulem},
  \citenamefont {Sulem},\ and\ \citenamefont {Frisch}}]{Sulem1983}%
  \BibitemOpen
  \bibfield  {author} {\bibinfo {author} {\bibfnamefont {C.}~\bibnamefont
  {Sulem}}, \bibinfo {author} {\bibfnamefont {P.-L.}\ \bibnamefont {Sulem}}, \
  and\ \bibinfo {author} {\bibfnamefont {H.}~\bibnamefont {Frisch}},\ }\href
  {\doibase 10.1016/0021-9991(83)90045-1} {\bibfield  {journal} {\bibinfo
  {journal} {Journal of Computational Physics}\ }\textbf {\bibinfo {volume}
  {50}},\ \bibinfo {pages} {138} (\bibinfo {year} {1983})}\BibitemShut
  {NoStop}%
\bibitem [{Note2()}]{Note2}%
  \BibitemOpen
  \bibinfo {note} {See Eq.~(2.2) of \cite {Sulem1983} for more
  details.}\BibitemShut {Stop}%
\bibitem [{Note3()}]{Note3}%
  \BibitemOpen
  \bibinfo {note} {We are grateful to P.~Bizon, M.~Maliborski, and
  A.~Rostworowski for suggesting we look at how higher derivatives
  behave.}\BibitemShut {Stop}%
\bibitem [{\citenamefont {Bizo\'{n}}\ and\ \citenamefont
  {Rostworowski}(2015)}]{Bizon:2014bya}%
  \BibitemOpen
  \bibfield  {author} {\bibinfo {author} {\bibfnamefont {P.}~\bibnamefont
  {Bizo\'{n}}}\ and\ \bibinfo {author} {\bibfnamefont {A.}~\bibnamefont
  {Rostworowski}},\ }\href {\doibase 10.1103/PhysRevLett.115.049101} {\bibfield
   {journal} {\bibinfo  {journal} {Phys. Rev. Lett.}\ }\textbf {\bibinfo
  {volume} {115}},\ \bibinfo {pages} {049101} (\bibinfo {year} {2015})},\
  \Eprint {http://arxiv.org/abs/1410.2631} {arXiv:1410.2631 [gr-qc]}
  \BibitemShut {NoStop}%
\bibitem [{\citenamefont {Balasubramanian}\ \emph {et~al.}(2015)\citenamefont
  {Balasubramanian}, \citenamefont {Buchel}, \citenamefont {Green},
  \citenamefont {Lehner},\ and\ \citenamefont
  {Liebling}}]{Balasubramanian:2015uua}%
  \BibitemOpen
  \bibfield  {author} {\bibinfo {author} {\bibfnamefont {V.}~\bibnamefont
  {Balasubramanian}}, \bibinfo {author} {\bibfnamefont {A.}~\bibnamefont
  {Buchel}}, \bibinfo {author} {\bibfnamefont {S.~R.}\ \bibnamefont {Green}},
  \bibinfo {author} {\bibfnamefont {L.}~\bibnamefont {Lehner}}, \ and\ \bibinfo
  {author} {\bibfnamefont {S.~L.}\ \bibnamefont {Liebling}},\ }\href {\doibase
  10.1103/PhysRevLett.115.049102} {\bibfield  {journal} {\bibinfo  {journal}
  {Phys. Rev. Lett.}\ }\textbf {\bibinfo {volume} {115}},\ \bibinfo {pages}
  {049102} (\bibinfo {year} {2015})},\ \Eprint
  {http://arxiv.org/abs/1506.07907} {arXiv:1506.07907 [gr-qc]} \BibitemShut
  {NoStop}%
\bibitem [{Note4()}]{Note4}%
  \BibitemOpen
  \bibinfo {note} {Here we use $t$ instead of $\tau $ for easier comparison to
  the literature.}\BibitemShut {Stop}%
\bibitem [{\citenamefont {Dimitrakopoulos}\ and\ \citenamefont
  {Yang}(2015)}]{Dimitrakopoulos:2015pwa}%
  \BibitemOpen
  \bibfield  {author} {\bibinfo {author} {\bibfnamefont {F.}~\bibnamefont
  {Dimitrakopoulos}}\ and\ \bibinfo {author} {\bibfnamefont {I.-S.}\
  \bibnamefont {Yang}},\ }\href {\doibase 10.1103/PhysRevD.92.083013}
  {\bibfield  {journal} {\bibinfo  {journal} {Phys. Rev.}\ }\textbf {\bibinfo
  {volume} {D92}},\ \bibinfo {pages} {083013} (\bibinfo {year} {2015})},\
  \Eprint {http://arxiv.org/abs/1507.02684} {arXiv:1507.02684 [hep-th]}
  \BibitemShut {NoStop}%
\bibitem [{\citenamefont {Ja\l{}mu\.zna}\ \emph {et~al.}(2011)\citenamefont
  {Ja\l{}mu\.zna}, \citenamefont {Rostworowski},\ and\ \citenamefont
  {Bizo\'{n}}}]{Jalmuzna:2011qw}%
  \BibitemOpen
  \bibfield  {author} {\bibinfo {author} {\bibfnamefont {J.}~\bibnamefont
  {Ja\l{}mu\.zna}}, \bibinfo {author} {\bibfnamefont {A.}~\bibnamefont
  {Rostworowski}}, \ and\ \bibinfo {author} {\bibfnamefont {P.}~\bibnamefont
  {Bizo\'{n}}},\ }\href {\doibase 10.1103/PhysRevD.84.085021} {\bibfield
  {journal} {\bibinfo  {journal} {Phys. Rev.}\ }\textbf {\bibinfo {volume}
  {D84}},\ \bibinfo {pages} {085021} (\bibinfo {year} {2011})},\ \Eprint
  {http://arxiv.org/abs/1108.4539} {arXiv:1108.4539 [gr-qc]} \BibitemShut
  {NoStop}%
\bibitem [{\citenamefont {Buchel}\ \emph {et~al.}(2012)\citenamefont {Buchel},
  \citenamefont {Lehner},\ and\ \citenamefont {Liebling}}]{Buchel:2012uh}%
  \BibitemOpen
  \bibfield  {author} {\bibinfo {author} {\bibfnamefont {A.}~\bibnamefont
  {Buchel}}, \bibinfo {author} {\bibfnamefont {L.}~\bibnamefont {Lehner}}, \
  and\ \bibinfo {author} {\bibfnamefont {S.~L.}\ \bibnamefont {Liebling}},\
  }\href {\doibase 10.1103/PhysRevD.86.123011} {\bibfield  {journal} {\bibinfo
  {journal} {Phys. Rev.}\ }\textbf {\bibinfo {volume} {D86}},\ \bibinfo {pages}
  {123011} (\bibinfo {year} {2012})},\ \Eprint {http://arxiv.org/abs/1210.0890}
  {arXiv:1210.0890 [gr-qc]} \BibitemShut {NoStop}%
\bibitem [{\citenamefont {Buchel}\ \emph {et~al.}(2013)\citenamefont {Buchel},
  \citenamefont {Liebling},\ and\ \citenamefont {Lehner}}]{Buchel:2013uba}%
  \BibitemOpen
  \bibfield  {author} {\bibinfo {author} {\bibfnamefont {A.}~\bibnamefont
  {Buchel}}, \bibinfo {author} {\bibfnamefont {S.~L.}\ \bibnamefont
  {Liebling}}, \ and\ \bibinfo {author} {\bibfnamefont {L.}~\bibnamefont
  {Lehner}},\ }\href {\doibase 10.1103/PhysRevD.87.123006} {\bibfield
  {journal} {\bibinfo  {journal} {Phys. Rev.}\ }\textbf {\bibinfo {volume}
  {D87}},\ \bibinfo {pages} {123006} (\bibinfo {year} {2013})},\ \Eprint
  {http://arxiv.org/abs/1304.4166} {arXiv:1304.4166 [gr-qc]} \BibitemShut
  {NoStop}%
\bibitem [{\citenamefont {Moschidis}(2017)}]{Moschidis:2017llu}%
  \BibitemOpen
  \bibfield  {author} {\bibinfo {author} {\bibfnamefont {G.}~\bibnamefont
  {Moschidis}},\ }\href@noop {} {\  (\bibinfo {year} {2017})},\ \Eprint
  {http://arxiv.org/abs/1704.08681} {arXiv:1704.08681 [gr-qc]} \BibitemShut
  {NoStop}%
\bibitem [{\citenamefont {Moschidis}(2018)}]{Moschidis:2018ruk}%
  \BibitemOpen
  \bibfield  {author} {\bibinfo {author} {\bibfnamefont {G.}~\bibnamefont
  {Moschidis}},\ }\href@noop {} {\  (\bibinfo {year} {2018})},\ \Eprint
  {http://arxiv.org/abs/1812.04268} {arXiv:1812.04268 [math.AP]} \BibitemShut
  {NoStop}%
\bibitem [{\citenamefont {Dias}\ and\ \citenamefont
  {Santos}(2016)}]{Dias:2016ewl}%
  \BibitemOpen
  \bibfield  {author} {\bibinfo {author} {\bibfnamefont {O.~J.~C.}\
  \bibnamefont {Dias}}\ and\ \bibinfo {author} {\bibfnamefont {J.~E.}\
  \bibnamefont {Santos}},\ }\href@noop {} {\  (\bibinfo {year} {2016})},\
  \Eprint {http://arxiv.org/abs/1602.03890} {arXiv:1602.03890 [hep-th]}
  \BibitemShut {NoStop}%
\end{thebibliography}%
\end{document}